\documentclass[nofootinbib,nobibnotes]{revtex4}
\usepackage{epsfig}
\usepackage{dcolumn}
\usepackage{bm}

\usepackage{graphicx}
\usepackage{latexsym}

\def\be{\begin{equation}}
\def\ee{\end{equation}}
\def\bea{\begin{eqnarray}}
\def\eea{\end{eqnarray}}

\begin{document}

\title{Dark Energy and Neutrino Mass Limits from Baryogenesis}
\author{Peihong Gu}
\email{Guph@mail.ihep.ac.cn}
\author{Xiulian Wang}
\email{Wangxl@mail.ihep.ac.cn}
\author{Xinmin Zhang}
\email{xmzhang@mail.ihep.ac.cn} \affiliation{Institute of High
Energy Physics, Chinese Academy of Sciences, P.O. Box 918-4,
Beijing 100039, People's Republic of China}

\begin{abstract}
In this brief report we consider couplings of the dark energy
scalar, such as Quintessence to the neutrinos and discuss its
implications in studies on the neutrino mass limits from
Baryogenesis. During the evolution of the dark energy scalar, the
neutrino masses vary, consequently the bounds on the neutrino
masses we have here differ from those obtained before.
\end{abstract}
 \maketitle

 There are strong evidences that the Universe is
 spatially accelerating at the present time\cite{Pelmutter}.
 The simplest account of this cosmic acceleration seems to
  be a remnant small cosmological constant, however many physicists
  are attracted by the idea that a new form of matter, usually
  called dark energy\cite{M.S.Turner} is causing the cosmic
  accelerating. A simple candidate for dark energy is
Quintessence\cite{B.Ratra,C.Wetterich,J.A.Frieman,I.Zlatev}, a
scalar
  field (or multi scalar
  fields) with
  a canonical kinetic term and a potential term in
  the Lagrangian.
Another one is called in the literature as
  k-essence\cite{C.Armendariz,T.Chiba}. Differing from
  Quintessence, for k-essence the accelerating expansion of the
  Universe is driven by its kinetic rather than potential energy.
  Being a dynamical component, the scalar field dark energy is
  expected to interact with the ordinary matters. In the literature there
have been a lot of studies
  on the possible couplings of quintessence to
  baryons, dark matter and
  photons\cite{S.M.Carroll,R.D.Peccei,L.Amendola,P.J.E.Peebles,R.Horvat}.
  For example, in Refs.\cite{L.Amendola,L.Amendolaetal} it is shown that introducing interaction
between Quintessence and dark matter provides a solution to the
puzzle why $\Omega_{DM}$ and $\Omega_{DE}$ are nearly equal today.
Specifically the authors of Ref.\cite{Comelli} recently considered
a model of interacting dark energy with dark matter and in their
scenario the mass of dark matter particle $\chi$
  depends exponentially on the dark energy field scalar
  $Q, M_{\chi}( Q )=\bar{M}e^{-\lambda Q/ M_{pl} }$.
  During the evolution of the dark energy scalar field the mass of
dark matter particle varies, consequently the parameters of the
dark matter model, such as the minimal supersymmetric standard
model (MSSM) differs drastically from
   the results where no connection between dark energy and dark
matter is
   present.
Recent data on
  the possible variation of the electromagnetic fine structure
  constant reported in\cite{J.K.Webb et al} has triggered
  interests in studies related to the interactions between
  Quintessence and the matter fields. For this case, one usually
introduces an interaction
 of form
 $ \sim Q F_{\mu\nu}F^{\mu\nu}$
with $F_{\mu\nu}$ being the electromagnetic field strength tensor.

In the recent years we \cite{limingzhe2} have studied the possible
interactions between the dark energy scalars, such as Quintessence
or K-essence, and the matter fields of the standard electroweak
theory and have shown that during the evolution of these scalar
fields CPT symmetry is violated and the baryon number asymmetry
required is generated. The mechanism for baryogenesis and/or
leptogenesis proposed in Refs.\cite{limingzhe2,trodden} provides a
unified picture for dark energy and baryon matter of our
   Universe. In this brief report we consider possible couplings of
Quintessence to the neutrinos and study its effects on the
neutrino mass limits from Baryogenesis.

We start with an examination on the neutrino mass limits required
by avoiding the washing out of the baryon number
asymmetry\cite{fy}. Consider a dimension 5 operator
  \be
  L_{\not L}=\frac{2}{f}l_{L}l_{L}\phi\phi+h.c,
  \ee
  where $f$ is a scale of new physics beyond the Standard Model
  which generates the $B-L$ violations, $l_{L},  \phi$ are the
  left-handed lepton and Higgs doublets respectively. When the
  Higgs field gets a vacuum expectation value $<\phi> \sim v$, the
  left-handed neutrino receives a majorana mass
  $m_{\nu} \sim \frac{v^{2}}{f}$. If this interaction in the early
universe is strong enough, combined with the electroweak Sphaleron
effect it will wash out any baryon number asymmetry of the
Universe.

At finite temperature,
the lepton number violating rate induced
  by the interaction in
  Eq.(1) is\cite{U.Sarkar}
\be
 \Gamma_{\not L}\sim 0.04\frac{T^{3}}{f^{2}} .
 \ee
  The survival
of the baryon number asymmetry requires
  this rate to be smaller than the Universe expansion rate
  $ H \sim 1.66g^{1/2}_{\ast}T^{2}/M_{pl}$,
which gives rise to a T-dependent
upper
  limit on the neutrino
  mass
\be
 \Sigma m_{\nu i}^{2} =
  [0.2eV(\frac{10^{12}GeV}{T})^{1/2}]^{2}.
  \ee
For instance, taking T around 100 GeV for  one type of neutrino it
gives $m_\nu < 20$KeV, however for $T \sim 10^{10}$GeV, a typical
leptogenesis temperature, this bound reduces to 2 eV.

Now we introduce an interaction between the neutrinos and the
Quintessence \be
 \beta \frac { Q }{M_{pl}} \frac{2 }{f} l_{L}l_{L}
\phi \phi+ h.c  , \ee
 where $\beta $ is the coefficient which charactrizes the strength of the
Quintessence interacting with the neutrinos and generally one
requires  $\beta < 4 \pi$ to make the effective lagrangian
description reliable. Combining Eq.(1) and Eq.(4) we have an
effective operator for  Quintessence-dependent neutrino masses \be
 L_{\not L}(Q)  =\frac{2 C(Q)}{f} l_{L}l_{L} \phi
\phi+h.c, \ee
 where $C(Q)= 1 + \beta Q/ M_{pl}$.

In the early universe the $B-L$ violating interaction rate now
becomes
  $$\Gamma_{\not L} \sim 0.04\frac{T^{3}}{f^{2}} C^{2}(Q). $$
Correspondingly the formula for the neutrino mass upper limit in
(3) changes to
 $$\sum m^{2}_{\nu i}=[0.2 {\rm eV} (\frac{10^{12}{\rm GeV}}{T })^{\frac{1}{2}}
\frac{C(Q_{0})}{C(Q_T )}]^{2}, $$
  where $Q_{0}$ is the value of the quintessence field at present time and
$Q_T$ the Quintessence evaluated at the temperature T. In general
$\frac{C(Q_{0})}{C(Q_T)}$ will not be one, so one expects a change
on the neutrino mass limit for a given temperature T.

To evaluate $C(Q)$ we need to solve the following equations of
motion of the Quintessence,  which for a flat Universe are given
by,
  \be
  H^2=\frac{8 \pi G}{3}(\rho_B +\frac{\dot{Q}^2}{2} +V(Q)),
  \ee
  \be
  \ddot{Q}+3H\dot{Q}+V'(Q)=0,
  \ee
  \be
  \dot{H}=-4\pi G ((1+w_B)\rho_B+\dot{Q}^2),
  \ee
where $\rho_B$ and $"w_B"$ represent respectively the energy
density and the equation-of-state of the background fluid, for
example $w_B=1/3$ in radiation-dominated and $w_B=0$ in the
matter-dominated Universe.

For numerical studies, we consider a model of Quintessence with a
inverse power-law potential\cite{B.Ratra},
 \be
 \label{modela}
 V=V_0 Q^{-\alpha}.
 \ee
This model is shown \cite{B.Ratra,I.Zlatev}to have the property of
tracking behavior. For a general discussion on the tracking
solution, one considers a function $\Gamma \equiv V''V/(V')^2$,
which when combining with Eqs.(6)(7), is given by
 \be \Gamma =1+\frac{w_B
-w_Q}{2(1+w_Q)}-\frac{1+w_B-2w_Q}{2(1+w_Q)}\frac{\dot{x}}
{6+x}-\frac{2}{(1+w_Q)}\frac{\ddot{x}}{(6+\dot{x})^2},
 \ee
 where $x \equiv (1+w_Q)/(1-w_Q), \dot{x} \equiv d \ln x/d \ln a$ and
 $\ddot{x} \equiv d^2 \ln x /d \ln a^2$. If $w_Q < w_B, \Gamma>1$
 and $\Gamma$ is nearly constant (i.e, $|d(\Gamma-1)/Hdt|
 <<|\Gamma-1|)$\cite{I.Zlatev}, the model has the tracking
 property.

\begin{figure}
\includegraphics[scale=.25]{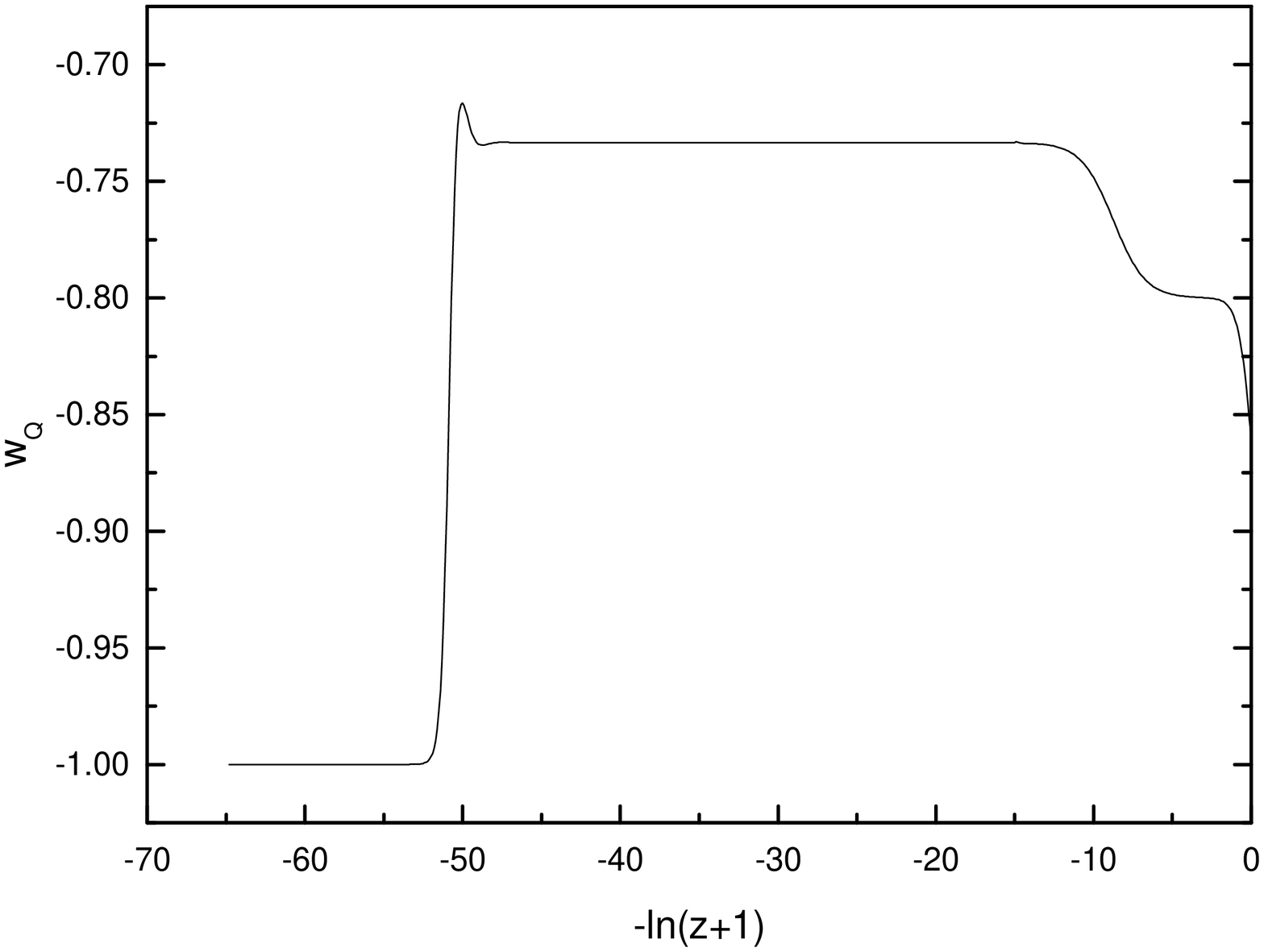}
\includegraphics[scale=.25]{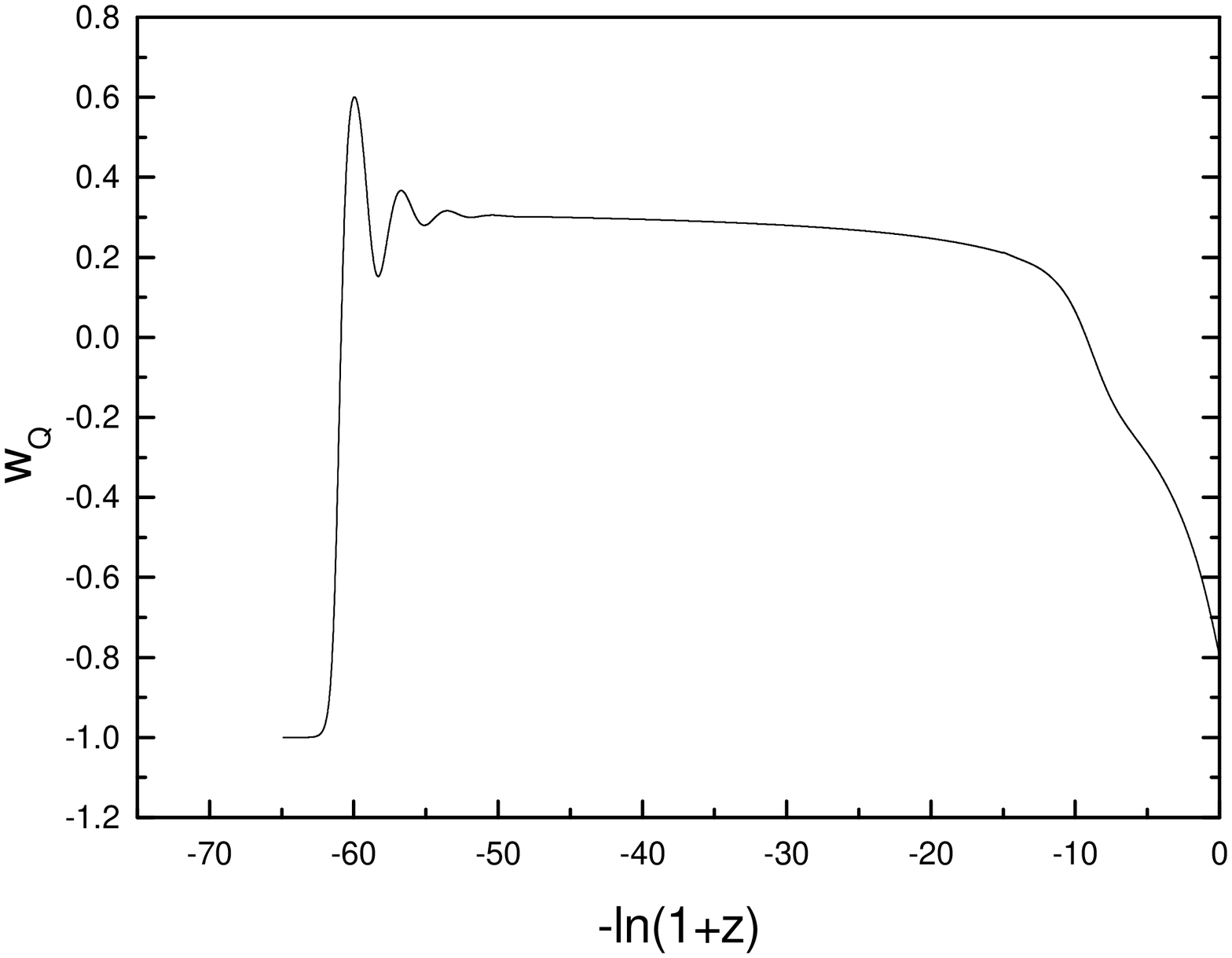}
\caption{Plot of $w_Q$ as a function of $-\ln(1+z)$. The left one
is for quintessence model given by Eq.(\ref{modela}); the right
one is for quintessence model given by Eq.(\ref{modelb}).
}\label{q05}
\end{figure}

 In the tracking region,
 \be
 \Gamma-1=\frac{w_B -w_Q}{2(1+w_Q)}=\frac{1}{\alpha},
 \ee
 then $w_Q =(\alpha w_B -2)/(\alpha +2)$. The WMAP gives that
 $w_{Q0}<-0.78$\cite{Spergel}, which requires a small value of $\alpha$ for this model.
 In Fig.\ref{q05}, we show the evolution of $w_Q$ with time,
 for parameters $\alpha =0.5$, $\Omega_{Q_0} \simeq
 0.7 $.
In Fig.2 we plot the evolution of Quintessence field as a function
of redshift z. The values of Quintessence field at present time
 $Q_{0}$ is $0.143M_{pl}$.
\begin{figure}
\vspace{2.5cm}
\includegraphics[scale=.25]{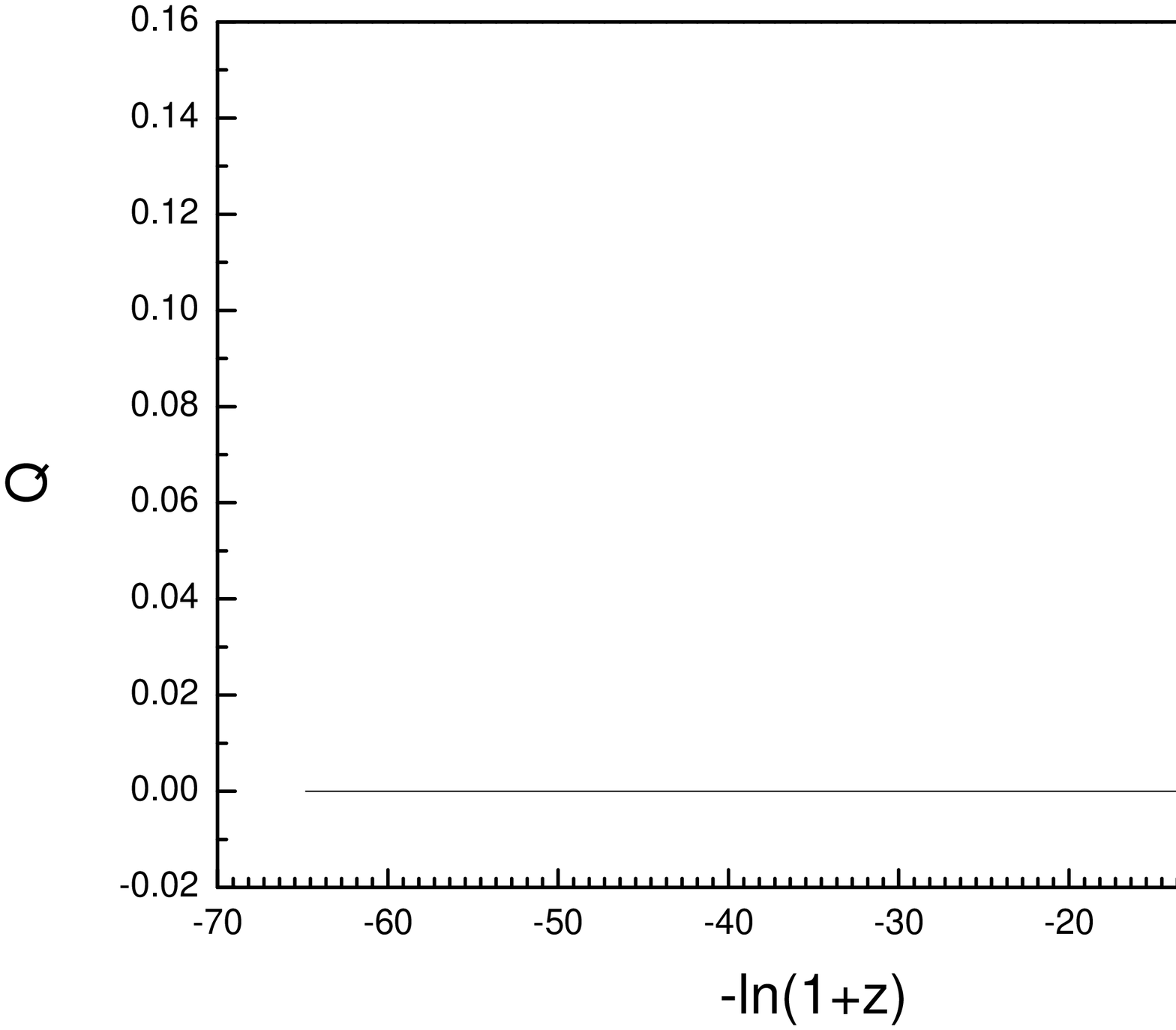}
\includegraphics[scale=.25]{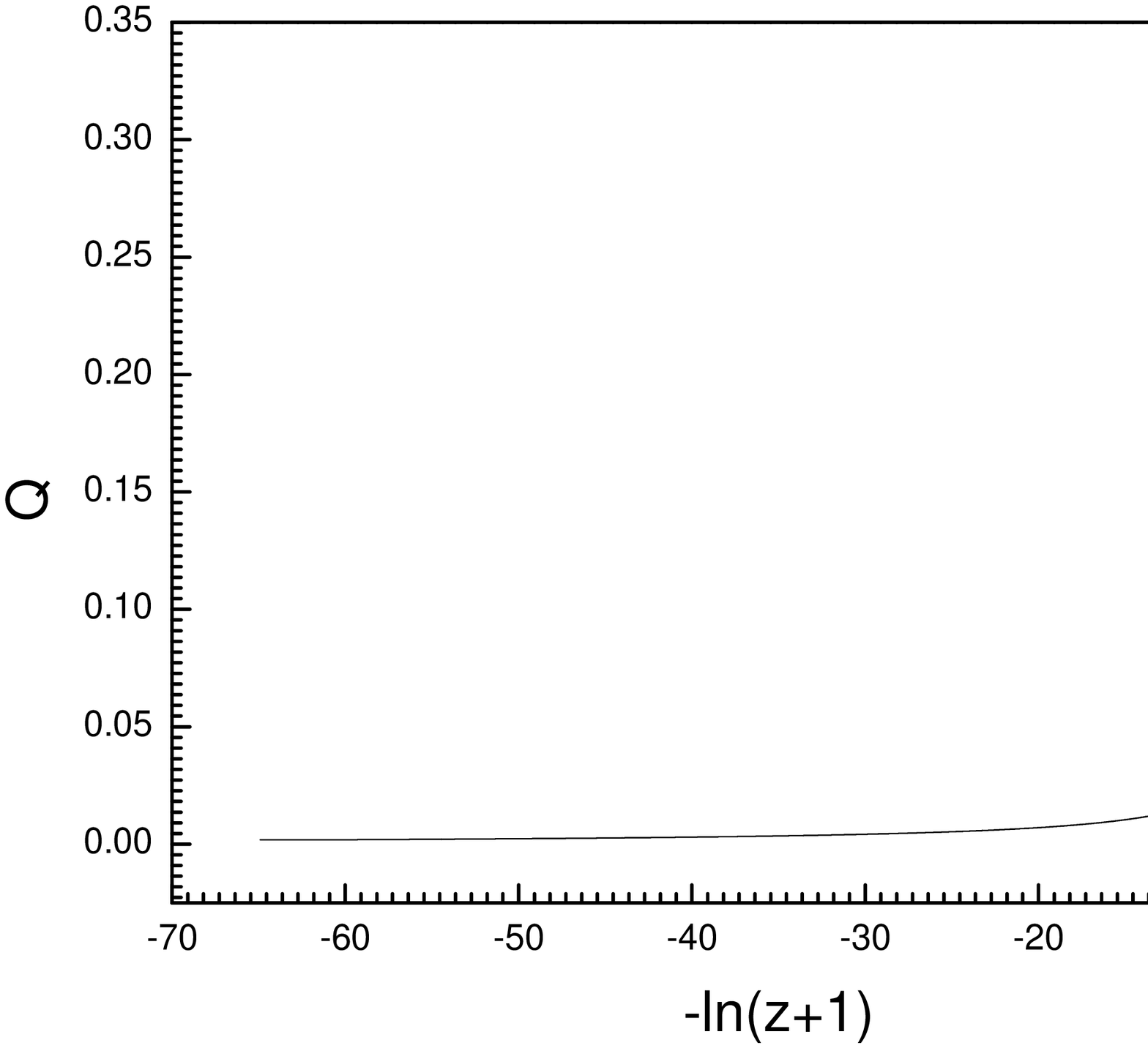}
\vspace{-2.5cm} \caption{Plot of $Q$ expressed in unit of $M_{pl}$
as a function of $-\ln(1+z)$. The left one is for quintessence
model given by Eq.(\ref{modela}); the right one is for
quintessence model given by Eq.(\ref{modelb}). }\label{q05qlnz}
\end{figure}


Defining $\Sigma \equiv (\Sigma m_{\nu i}^2)^{\frac{1}{2}}
=\frac{C(Q_0)}{C(Q_T)} \Sigma_T$ where $ \Sigma_T =0.2 eV
(\frac{10^{12}GeV}{T })^{1/2}$, we in Fig.3 plot the $\Sigma$ as a
function of the temperature T for different values of parameters
$\beta$. $\beta = 0$ corresponds to the case when no interaction
between the Quintessence and the neutrinos exist. One can see from
the figure that the difference between $\beta = 0$ and $\beta
\not= 0$ increases as the temperature decreases.
\begin{figure}
\includegraphics[scale=.3]{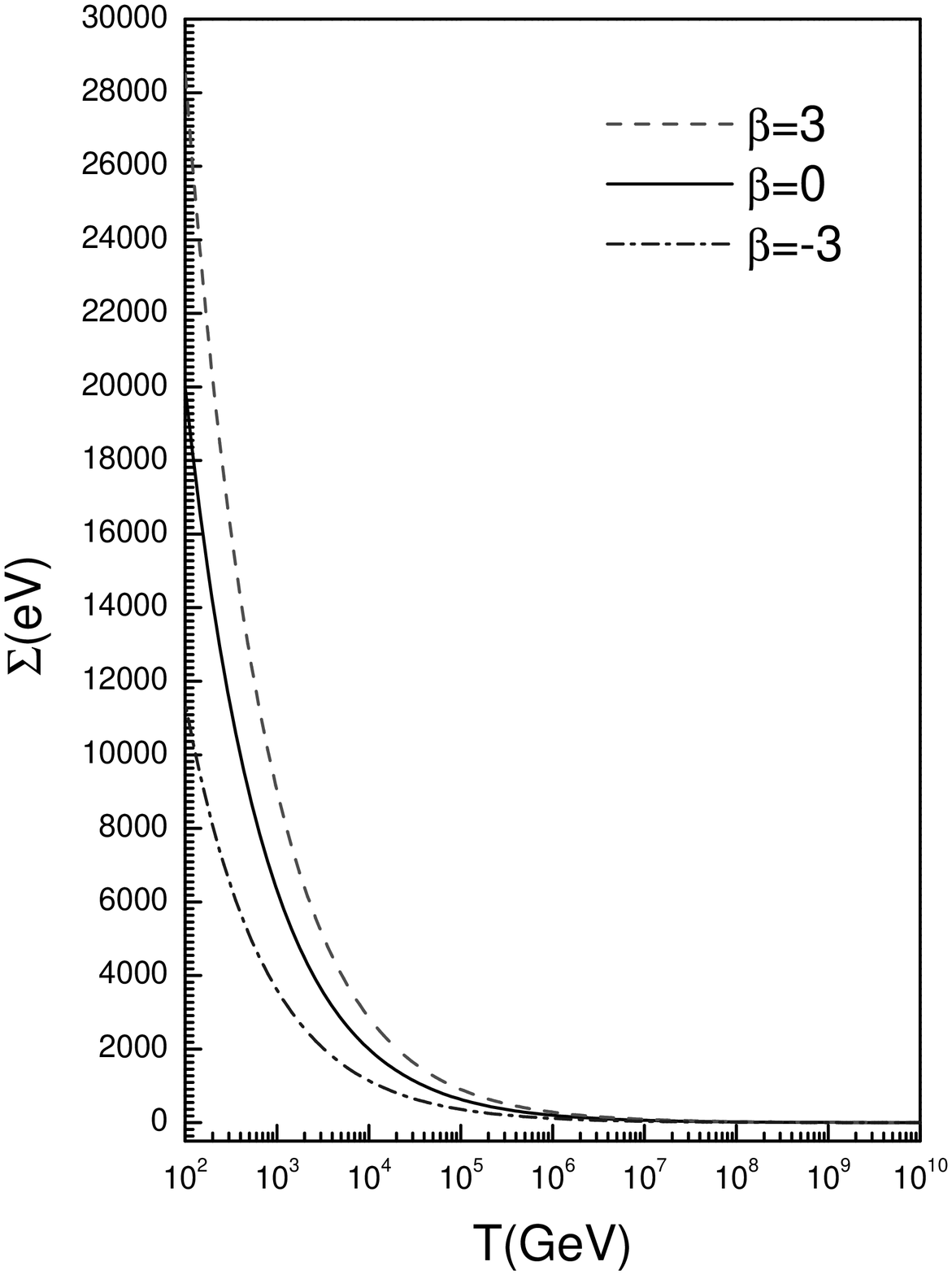}
\includegraphics[scale=.3]{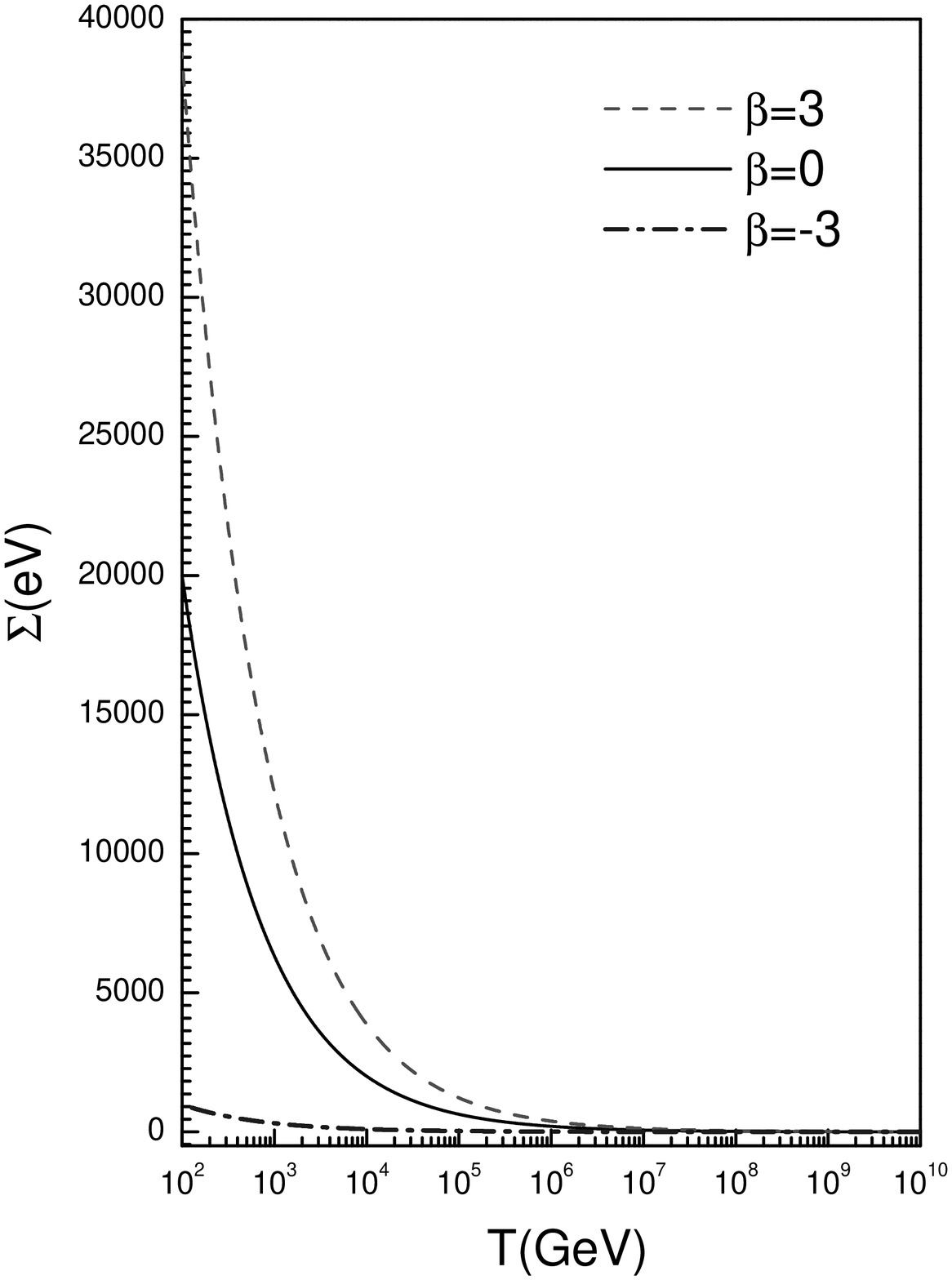}
\caption{Plot of $\sum$ for different values of $\beta$ as a
function of temperature $T$. The left one is for quintessence
model given by Eq.(\ref{modela}); the right one is for
quintessence model given by Eq.(\ref{modelb}). }\label{e05mpl}
\end{figure}
Numerically we find at $T \sim 100$ GeV, for one type of neutrino
the mass bound which is 20 KeV for $\beta = 0$ changes to 29 KeV
for $\beta = 3$ and 11 KeV for $\beta = -3$. At $T\sim 10^{10}$
GeV, these mass limits are 2.9 eV, 2 eV and 1.1 eV for $\beta = 3,
0, -3$ respectively.

Our limits on the neutrino masses depend on the Quintessence
model. For an illustration, we consider another Quintessence
model\cite{I.Zlatev}
 \be \label{modelb} V(Q)=V_0 \exp(\lambda /Q).
\ee In Fig.1 we show the evolution of $\omega_Q$ with time. We
take $\lambda =0.5 M_{pl}$ which gives rise to $\omega_Q \simeq
-0.8 $ at present time in consistent with the WMAP limit. The
behavior of Quintessence field for this model as shown in Fig.2 is
different from the model we studied above. Consequently the
neutrino mass limits will also be different.
 From Fig.3 one can see that for this model the neutrino mass limits
 differ drastically from that
obtained in the absence of the Quintessence interacting with
neutrinos. For example taking $T=100$ GeV, $\Sigma$ are 39, 20, 1
KeV for $\beta = 3, 0, -3$ and at $T=10^{10}$ GeV $\Sigma = 3.9,
2, 0.1$ eV respectively.

In summary we have considered in this note a scenario where
neutrino masses vary during the evolution of the Dark Energy
scalars, such as Quintessence and studied its implications in
Baryogenesis. We assume that the neutrino masses are from a
dimension 5 operator in Eq.(1) and the interaction form of the
Quintessence with the neutrinos is given by (4). The operator (1)
is not renormalizable, which in principle can be generated by
integrating out the heavy particles. For example, in the model of
the minimal see-saw mechanism \cite{seesaw}for the neutrino masses
\be
L=h_{ij}\bar{l}_{Li}N_{Rj}\phi+\frac{1}{2}M_{ij}\bar{N}^{c}_{Ri}N_{Rj}+h.c.
\ee
 where $ M_{ij}$ is the mass matrix of
the right-handed neutrinos and the Dirac mass of neutrino is given
by $m_{D}\equiv h_{ij} <\phi> $. Integrating out the heavy
right-handed neutrinos will generate the operator in (1), however
to have the light neutrino masses varied there are various
possibilities, such as by coupling the Quintessence field to
either the Dirac masses or the majorana masses of the right-handed
neutrinos or both. Qualitatively because of these interactions the
neutrino mass limits from leptogenesis\cite{fukugida} are expected
to be changed, however to quantify these changes one needs to
specify the details of these couplings and the Quintessence
models. Numerical studies on leptogenesis in the minimal see-saw
model show that the neutrino mass is bounded from above which for
three degenerated neutrinos is $m_\nu < 0.12$
eV\cite{WBuchmuller0302092}. Interaction of the Quintessence with
the neutrinos can change this upper bound. Note that the
cosmological limit on the neutrino mass from WMAP gives $m_\nu <
0.23$eV\cite{Spergel}. Interestingly a recent study on the
cosmological data showed a preference for neutrinos with
degenerated masses around 0.21eV\cite{allen}.

Our studies in this brief report can be generalized into models of
electroweak baryogenesis in the discussions of the constraints on
the model parameters, such as the Higgs boson mass.

{ \bf Acknowledgement:} This work is supported in part by Natural
Science Foundation of China and by Ministry of Science and
Technology of China under Grant No. NKBRSF G19990754.

\newcommand\AJ[3]{~Astron. J.{\bf ~#1}, #2~(#3)}
\newcommand\APJ[3]{~Astrophys. J.{\bf ~#1}, #2~ (#3)}
\newcommand\APJL[3]{~Astrophys. J. Lett. {\bf ~#1}, L#2~(#3)}
\newcommand\APP[3]{~Astropart. Phys. {\bf ~#1}, #2~(#3)}
\newcommand\CQG[3]{~Class. Quant. Grav.{\bf ~#1}, #2~(#3)}
\newcommand\JETPL[3]{~JETP. Lett.{\bf ~#1}, #2~(#3)}
\newcommand\MNRAS[3]{~Mon. Not. R. Astron. Soc.{\bf ~#1}, #2~(#3)}
\newcommand\MPLA[3]{~Mod. Phys. Lett. A{\bf ~#1}, #2~(#3)}
\newcommand\NAT[3]{~Nature{\bf ~#1}, #2~(#3)}
\newcommand\NPB[3]{~Nucl. Phys. B{\bf ~#1}, #2~(#3)}
\newcommand\PLB[3]{~Phys. Lett. B{\bf ~#1}, #2~(#3)}
\newcommand\PR[3]{~Phys. Rev.{\bf ~#1}, #2~(#3)}
\newcommand\PRL[3]{~Phys. Rev. Lett.{\bf ~#1}, #2~(#3)}
\newcommand\PRD[3]{~Phys. Rev. D{\bf ~#1}, #2~(#3)}
\newcommand\PROG[3]{~Prog. Theor. Phys.{\bf ~#1}, #2~(#3)}
\newcommand\PRPT[3]{~Phys.Rept.{\bf ~#1}, #2~(#3)}
\newcommand\RMP[3]{~Rev. Mod. Phys.{\bf ~#1}, #2~(#3)}
\newcommand\SCI[3]{~Science{\bf ~#1}, #2~(#3)}
\newcommand\SAL[3]{~Sov. Astron. Lett{\bf ~#1}, #2~(#3)}

\end{document}